\documentstyle[prd,preprint,aps]{revtex}

\tighten

\begin{document}

\preprint{SFU HEP-151-98}

\draft

\title{Tadpole renormalization and relativistic corrections
in lattice NRQCD}

\author{Norman H. Shakespeare
\footnote{Email address: nshakesp@sfu.ca.}
and Howard D. Trottier
\footnote{Email address: trottier@sfu.ca.}}
\address{Department of Physics, Simon Fraser University, 
Burnaby, B.C., Canada V5A 1S6}

\date{February 1998}

\maketitle

\begin{abstract}
\noindent
We make a detailed comparison of two tadpole renormalization 
schemes in the context of the quarkonium hyperfine splittings 
in lattice NRQCD. We renormalize improved gauge-field and
NRQCD actions using the mean-link $u_{0,L}$ in Landau gauge,
and using the fourth root of the average plaquette $u_{0,P}$.
Simulations are done for the three quarkonium systems $c\bar c$, 
$b\bar c$, and $b\bar b$. The hyperfine splittings 
are computed both at leading ($O(M_Q v^4)$) and at next-to-leading 
($O(M_Q v^6)$) order in the relativistic expansion, where $M_Q$ 
is the renormalized quark mass, and $v^2$ is the mean-squared velocity. 
Results are obtained at a large number of 
lattice spacings, in the range of about 0.14~fm to 0.38~fm.
A number of features emerge, all of which favor tadpole
renormalization using $u_{0,L}$. This includes much better scaling 
behavior of the hyperfine splittings in the three quarkonium systems 
when $u_{0,L}$ is used. We also find that relativistic corrections
to the spin splittings are smaller when $u_{0,L}$ is used,
particularly for the $c\bar c$ and $b\bar c$ systems.
We also see signs of a breakdown in the NRQCD expansion 
when the bare quark mass falls below about one in lattice units.
Simulations with $u_{0,L}$ also appear to be better behaved in this
context: the bare quark masses turn out to be larger when $u_{0,L}$ 
is used, compared to when $u_{0,P}$ is used on lattices with 
comparable spacings. These results also demonstrate the need to go 
beyond tree-level tadpole improvement for precision simulations.
\end{abstract}

\section{Introduction}
Tadpole improvement of lattice actions \cite{LepMac} has become an 
essential ingredient in numerical simulations of hadronic systems. 
Tadpole improvement has revitalized interest in simulations on coarse 
lattices, and is playing an important role in current efforts to extract 
continuum results from simulations on fine lattices.

Tadpole diagrams in lattice theories are induced by the nonlinear 
connection between the lattice link variables $U_\mu$ and the
continuum gauge fields. This has been shown to cause large radiative 
corrections to many quantities in lattice theories. 
Fortunately, most of the effects of tadpoles can be removed by a 
simple mean field renormalization of the links \cite{LepMac}
\begin{equation}
   U_\mu(x) \rightarrow {U_\mu(x) \over u_0} ,
\label{u0}
\end{equation}
where an operator dominated by short-distance fluctuations
is used to determine $u_0$. 

One of the earliest applications of tadpole improvement was in the 
development of lattice nonrelativistic quantum chromodynamics (NRQCD) 
\cite{LepThac,N1992,Nalphas,Nbmass,Nups,Npsi}. Precision simulations 
of the $\Upsilon$ system in NRQCD have provided many important 
phenomenological results, including the strong coupling constant
\cite{Nalphas} and the $b$-quark pole mass \cite{Nbmass,Nups}. 
However the situation for charmonium is more problematic, due to
large relativistic corrections in this system \cite{HDT}.

In fact the quarkonium spectrum provides a powerful probe of 
tadpole renormalization. The quarkonium fine and hyperfine 
spin splittings in particular are very sensitive to the details of 
the NRQCD Hamiltonian, with the relevant operators undergoing large 
tadpole renormalizations. For example, it has been shown \cite{HDT} 
that scaling of the charmonium hyperfine splitting is significantly 
improved when the tadpole renormalization is determined using 
the mean-link $u_{0,L}$ measured in Landau gauge \cite{LepMac}:
\begin{equation}
   u_{0,L} \equiv
       \left\langle \case13 \mbox{ReTr} \, U_\mu \right\rangle, 
       \quad \partial_\mu A_\mu = 0 ,
\label{ulandau}
\end{equation}
compared to when the fourth root of the average plaquette $u_{0,P}$
is used:
\begin{equation}
   u_{0,P} \equiv
       \left\langle \case13 \mbox{ReTr} \, U_{\mbox{pl}} 
       \right\rangle^{1/4} .
\label{uplaq}
\end{equation}
Equation (\ref{uplaq}) has been employed in most previous 
lattice simulations. First evidence for a more continuum-like 
behavior of lattice actions using $u_{0,L}$ came from studies of 
rotational symmetry restoration in the heavy quark potential 
\cite{LepagePot}. More recently, improved scaling of the charmonium 
spectrum from relativistic actions \cite{D234}, and of the SU(2) 
glueball spectrum \cite{NSglueball}, have also been observed in 
simulations using $u_{0,L}$.

In this paper we make a detailed comparison of the two tadpole
renormalization schemes $u_{0,L}$ and $u_{0,P}$ (when implemented 
at tree-level) in the context of the quarkonium hyperfine splittings 
in NRQCD. This is done for the three quarkonium systems 
$c\bar c$, $b\bar c$, and $b\bar b$. The hyperfine splittings 
are computed both at leading ($O(M_Q v^4)$) and at next-to-leading 
($O(M_Q v^6)$) order in the relativistic expansion, 
where $M_Q$ is the renormalized quark mass, and $v^2$ is the 
mean-squared velocity. Results are obtained at a large number of 
lattice spacings, in the range of about 0.14~fm to 0.38~fm. All 
quantities are calculated in the two tadpole schemes after careful 
re-tuning of the lattice action parameters for each system.

We find that scaling of the hyperfine splittings in all three
quarkonium systems is significantly improved when $u_{0,L}$ is used.
This confirms and extends the scaling analysis done in Ref. \cite{HDT}.
In addition we see signs of a possible breakdown of the NRQCD 
effective action at smaller lattice 
spacings, when the bare quark mass in lattice units $aM_Q^0$ 
falls below about one (scaling studies of the $\Upsilon$ spectrum, as a 
probe of the NRQCD expansion, have previously been reported in 
Ref.\ \cite{bbscaling}). Large changes in the $c\bar c$ and $b\bar c$
splittings with $u_{0,P}$ are evident at smaller lattices spacings.
Once again, simulations with $u_{0,L}$ are better behaved
in this context: the 
bare charm-quark mass turns out to be much larger when $u_{0,L}$ 
is used, compared to when $u_{0,P}$ is used on lattices with 
comparable spacings. Although there are clear pathologies in the 
$u_{0,P}$ data at small bare masses, our conclusions here are
necessarily tentative, since we are unable to do realistic simulations 
on sufficiently fine lattices to get $aM_c^0$ appreciably below one 
with $u_{0,L}$ tadpole improvement.

We also find that the choice of tadpole renormalization scheme 
has a strong effect on the apparent convergence of the 
velocity expansion which underlies the NRQCD effective action.
We find that the relativistic corrections to the hyperfine 
splittings are smaller when $u_{0,L}$ is used, particularly
for the $c\bar c$ and $b\bar c$ systems.
This casts new light on the results obtained in Ref.\ \cite{HDT},
where relativistic corrections to spin splittings in NRQCD were
first calculated. Results obtained here show that the charmonium 
hyperfine splitting is reduced by about 30\% when $u_{0,L}$
is used to renormalize the action. This is consistent with 
a naive estimate of $v^2_{c\bar c}$ from operator expectation 
values in NRQCD \cite{N1992}. Hence while the velocity expansion for 
charmonium is subject to large corrections, it may not be as 
unreliable as was suggested in Ref.\ \cite{HDT}, where 
simulations with $u_{0,P}$ on relatively fine lattices showed 
relativistic corrections of about 60\% in the charmonium hyperfine 
splitting. (Relativistic corrections have more recently been analyzed 
in the $\Upsilon$ system \cite{Manke,Spitz}, and have also
been studied in heavy-light mesons \cite{ArifaHL,Ishikawa,Lewis}).

Although these results clearly favor using $u_{0,L}$ for tadpole
improvement, they also serve to underline the need to go beyond
tree-level matching of lattice actions. Results from this study
and others (see e.g.\ Refs.\ 
\cite{LepagePot,D234,NSglueball,bbscaling}) demonstrate 
that precision results can only be obtained once uncertainties due 
to $O(\alpha_s)$ renormalizations are removed (for some recent work 
in this connection see e.g. Refs.\ \cite{Luscher,LepHDT}).

\section{Details of the Simulations}

\subsection{Lattice Actions}

The lattice NRQCD effective action for quarkonium is organized 
according to an expansion in the mean squared velocity 
$v^2$ of the heavy quarks, with corrections included 
for lattice artifacts. The effective action, including
spin-independent operators to $O(v^4)$, and spin-dependent 
interactions to $O(v^6)$, was derived in Ref.\ \cite{N1992}.
Following Refs.\ \cite{Nups,Npsi}, we use the evolution equation
\begin{equation}
   G_{t+1} =
   \left(1\!-\!\frac{aH_0}{2n}\right)^n
   U^\dagger_4
   \left(1\!-\!\frac{aH_0}{2n}\right)^n
   \left(1\!-\!a\delta H\right) G_t
   \quad (t>1) ,
\label{Gtp1}
\end{equation}
where the initial evolution is set by
\begin{equation}
   G_1 =
   \left(1\!-\!\frac{aH_0}{2n}\right)^n
   U^\dagger_4
   \left(1\!-\!\frac{aH_0}{2n}\right)^n \, \delta_{\vec x,0} .
\label{G1}
\end{equation}
On the lattice the leading kinetic energy operator $H_0$ is given by
\begin{equation}
   H_0 = - { \Delta^{(2)} \over 2M_Q^0 } ,
\label{H0}
\end{equation}
where $M_Q^0$ is the bare quark mass and $\Delta^{(2)}$
is the lattice Laplacian. 

Relativistic corrections are organized in powers of the heavy 
quark velocity:
\begin{equation}
   \delta H = \delta H^{(4)} + \delta H^{(6)} .
\label{deltaH}
\end{equation}
$\delta H^{(4)}$ contains spin-independent relativistic 
corrections and leading-order spin interactions:
\begin{eqnarray}
   \delta H^{(4)} 
 & = &
   - c_1 { ( \Delta^{(2)} )^2 \over 8(M_Q^0)^3 }
   + c_2 { ig \over 8 (M_Q^0)^2 }
         ( \tilde {\bf \Delta} \cdot \tilde{\bf E}
         - \tilde {\bf E} \cdot \tilde {\bf \Delta} )
\nonumber \\
  & & 
   - c_3 { g \over 8( M_Q^0 )^2 } 
         \mbox{{\boldmath$\sigma$}} \cdot 
         ( \tilde {\bf \Delta} \times \tilde {\bf E} 
         - \tilde {\bf E} \times \tilde {\bf \Delta} )
   - c_4 { g \over 2 M_Q^0 } 
         \mbox{{\boldmath$\sigma$}} \cdot \tilde {\bf B}
\nonumber \\
  & &  
   + c_5 { a^2 \Delta^{(4)} \over 24 M_Q^0 }
   - c_6 { a ( \Delta^{(2)} )^2 \over 16n (M_Q^0)^2 } ,
\label{H4}
\end{eqnarray}
with the last two terms coming from finite lattice spacing 
corrections to the lattice Laplacian and the lattice time derivative 
respectively. The parameter $n$ is introduced to remove instabilities 
in the heavy quark propagator caused by high momentum modes \cite{N1992}.
Spin-dependent relativistic corrections for quarkonium first
appear at $O(v^6)$:
\begin{eqnarray}
   \delta H^{(6)}
 & = &
   - c_7 { g \over 8 (M_Q^0)^3 } 
         \left\{  \tilde \Delta^{(2)} , 
                  \mbox{{\boldmath$\sigma$}} \cdot \tilde {\bf B} 
         \right\} ,
\nonumber \\
  & &
   - c_8 { 3g \over 64 (M_Q^0)^4 } 
         \left\{  \tilde \Delta^{(2)} , 
                  \mbox{{\boldmath$\sigma$}} \cdot 
                ( \tilde {\bf \Delta} \times \tilde {\bf E} 
                - \tilde {\bf E} \times \tilde {\bf \Delta} )
         \right\} ,
\nonumber \\
  & &
   - c_9 { i g^2 \over 8 (M_Q^0)^3 } 
         \mbox{{\boldmath$\sigma$}} \cdot 
         \tilde {\bf E} \times \tilde {\bf E} .
\label{H6}
\end{eqnarray}
Spin-independent corrections at $O(v^6)$ are not considered here
(these operators may in fact have indirect effects on spin splittings 
\cite{Npsi,Lewis}). As in Ref.\ \cite{HDT} simulations were done with 
the derivative operators and the clover fields corrected for their
leading discretization errors. This is indicated by the tilda 
superscripts on these operators in Eqs.\ (\ref{H4}) and (\ref{H6}).
Complete expressions for the operators can be found in
Refs. \cite{N1992,HDT}.

At tree-level all of the coefficients $c_i$ in Eqs.\ (\ref{H4}) and 
(\ref{H6}) are one. However tadpole improvement is crucial in order 
to eliminate large radiative corrections. Independent sets of 
simulations were done using the two tadpole renormalization 
schemes described in Section~I: the mean-link in Landau gauge, 
Eq.\ (\ref{ulandau}) (using a standard lattice implementation of 
the continuum Landau gauge fixing \cite{FFTLandau}), and the
fourth root of the average plaquette, Eq.\ (\ref{uplaq}).
The links were rescaled in the simulation before they were
input to the quark propagator subroutine, to be sure that
Eq. (\ref{u0}) was correctly implemented in all terms
in the heavy quark action. 
The gauge-field configurations were generated using an 
$O(a^4)$-accurate tadpole-improved action \cite{LepCoarse}
\begin{equation}
    S[U] = \beta \sum_{\mbox{pl}} \case13
           \mbox{ReTr} \left(1 - U_{\mbox{pl}}\right)
         - {\beta \over 20 u_0^2} \sum_{\mbox{rt}} \case13
           \mbox{ReTr} \left(1 - U_{\mbox{rt}}\right) ,
\label{Sglue}
\end{equation}
where the sums are over all oriented $1\times1$ plaquettes (pl)
and $1\times2$ rectangles (rt).

Meson creation operators were constructed from quark ($\psi^\dagger$)
and antiquark ($\chi^\dagger$) creation operators
\cite{LepThac,Nups,Npsi}:
\begin{equation}
   \sum_{\vec x} \psi^\dagger(\vec x) \Gamma(\vec x) 
                 \chi^\dagger(\vec x) ,
\end{equation}
using a gauge-invariant smearing function \cite{NoteSmear}
\begin{equation}
   \Gamma(\vec x) \equiv 
   \gamma^\dagger(\vec x) \Omega(\vec x) \gamma(\vec x) ,
\label{Gamma}
\end{equation}
where the $2 \times 2$ spin matrix $\Omega(\vec x)$ selects the 
quantum numbers of interest, and 
\begin{equation}
   \gamma(x) = \left( 1 + \epsilon \Delta^{(2)}(x) \right)^{n_s} .
\label{gamma}
\end{equation}
The weight $\epsilon$ and the number of smearing iterations $n_s$
were adjusted to optimize the overlap with the ground state.

Meson correlation functions \cite{Nups,Npsi} were computed for the 
${}^1S_0$ ($\Omega = I$), 
${}^3S_1$ ($\Omega = \sigma_i$) and 
${}^1P_1$ ($\Omega = \Delta_i$) mesons. 
The three triplet $P$-wave correlators 
(${}^3P_0$, ${}^3P_1$, ${}^3P_2$) were also analyzed, but
this data is of insufficient quality to report here.
Correlation functions were evaluated at 
both zero momentum and at the smallest allowed nonzero momentum.

\subsection{Simulation Parameters}
Six lattices were generated using the mean link in Landau
gauge to set the tadpole factor ($u_{0,L}$) and seven lattices with 
comparable spacings were generated using average plaquette 
tadpoles ($u_{0,P}$). The parameters of these thirteen lattices are 
given in Tables \ref{betaL} and \ref{betaP}. Note that we use
$\beta_L$ to denote the lattice coupling for simulations
with $u_{0,L}$, and $\beta_P$ for lattices with $u_{0,P}$.
To check that finite volume effects are not an issue
on the lattices with the smallest spacings, we performed
some runs at $\beta_P=7.3$ on a $16^4$ volume, and found 
no significant change in the results.

A standard Cabbibo-Marinari pseudo heat bath was used to generate
the gauge field configurations. The number of updates between 
measurements varied from 10 for lattices with the largest spacings 
to 20 for the smallest; autocorrelation times satisfied 
$\tau \alt 0.5$ in all cases. Smeared-smeared correlators were used, 
with typically 5--10 smearing iterations, and a smearing weight 
$\epsilon = 1/12$ was used in all cases.

The lattice spacings were determined from the spin-averaged
$1P - 1S$ mass difference, which we set to 458~MeV (the experimental 
value for charmonium); this mass difference is thought to be 
about the same for all quarkonium systems \cite{Nbc}. The 
difference between the singlet ${}^1P_1$ and the spin-averaged 
${}^3S_1$, ${}^1S_0$ states was used for this purpose.

The bare quark masses were tuned by calculating the kinetic
masses $M_{\rm kin}$ of the ${}^1S_0$ states in physical units,
which were extracted from fits to the energies $E_{\bf P}$ 
of the boosted states using the form
\begin{equation}
   E_{\bf P} - E_0 = { {\bf P}^2 \over 2 M_{\rm kin} } .
\label{Ep}
\end{equation}
Fits were made to the state with momentum components $(1,0,0)$
in units of $2\pi/(Na)$; relativistic corrections to the
dispersion relation made little difference in the fit values
of $M_{\rm kin}$. The bare quark masses $M_Q^0$ were fit
to the following ${}^1S_0$ kinetic masses:
$M_{c\bar c} = 2.98$~GeV, 
$M_{b\bar c} = 6.28$~GeV, and
$M_{b\bar b} = 9.46$~GeV.
The value of the $b \bar c$ $S$-wave meson mass adopted here was 
obtained in a previous NRQCD analysis of the 
$b \bar c$ system \cite{Nbc}.

A potential complication arises in that the lattice 
spacing and quark masses 
determined from different systems on the same quenched configurations 
do not generally agree. Since we compare results for a given system 
obtained over a wide range of lattice spacings, we have attempted to
minimize systematic effects from tuning errors by re-tuning the 
parameters for each system on each ensemble of configurations.
In the case of the $b\bar c$ system we used the $c$-quark mass
determined from the $M_{c\bar c}$ kinetic mass, and tuned the 
$b$-quark mass to reproduce $M_{b\bar c}$. Note that the $b$-quark
mass was re-tuned in simulations of the $b \bar b$ system, to 
obtain the correct value of $M_{b\bar b}$.
The final simulation results for the kinetic masses are 
accurate to within 3\% in all cases.

The resulting quark masses and lattice spacings for
the three quarkonium systems for the NRQCD action
at $O(v^6)$ are given in Tables \ref{TbaremL} and \ref{TbaremP}.
Note again that the lattice spacing is given separately for each system, 
the differences possibly reflecting effects due to quenching,
which has been conjectured to play a role in setting the scale 
in these and other hadronic systems (see e.g. Ref. \cite{Nbc}).
The $b$-quark mass is similarly given separately for the 
$b\bar c$ and $b\bar b$ systems. We found only small changes in 
the quark masses and lattice spacings when they are determined 
from the NRQCD action at $O(v^4)$.

Although we think that it is worthwhile to minimize systematic
errors by re-tuning the quenched lattice parameters for the
different quarkonium systems, and have done so in all
simulations reported here, these effects are actually small.
It is important to note that none of the conclusions reached in this 
study are changed if our re-tuning procedure is modified.

\section{Results and Analysis}

Hyperfine splittings for the three quarkonium systems
$c\bar c$, $b \bar c$, and $b \bar b$ were extracted on
the six lattices with $u_{0,L}$ (cf.\ Table \ref{betaL}),
and the seven lattices with $u_{0,P}$ (cf.\ Table \ref{betaP}).
To illustrate the quality of the raw data we show
effective mass plots $m_{\rm eff}(T) = -\log(G(T)/G(T-1))$ for 
the $b \bar c$ system at $O(v^6)$, on the two lattices with the 
smallest spacings, in Fig.\ \ref{FbcL} ($\beta_L=7.5$) 
and Fig.\ \ref{FbcP} ($\beta_P=7.3$).

Single exponential fits to the correlation functions were used to 
get the masses of the ground states, and a jackknife analysis
was used to estimate statistical errors. The hyperfine splittings were
obtained from a fit to the ratio of ${}^3S_1$ and ${}^1S_0$
correlation functions. Detailed fit results corresponding to the 
data in Figs. \ref{FbcL} and \ref{FbcP} are given in 
Tables \ref{TbcL} and \ref{TbcP}. Final estimates of the 
dimensionless energies were obtained by finding two or three 
successive $t_{\rm min}/t_{\rm max}$ intervals for which the fit 
results overlap within statistical errors. The largest statistical
error in the overlapping fits was used as an estimate of the 
error in the final fit value. We give the final fit
results for the dimensionless energies for the $b\bar c$ states 
at $O(v^6)$ in Tables~\ref{TbcfitL} and \ref{TbcfitP}.

The final fit results for all hyperfine splittings in physical
units are given in Tables~\ref{THL} and \ref{THP}, where the quoted
errors are purely statistical. We expect
a systematic error of order 10\% in the hyperfine splittings, coming 
from uncertainties in the quark mass determinations due to $O(v^6)$ 
spin-independent relativistic corrections \cite{Npsi,Lewis}.
Effects due to quenching may also be large in the case of
the charmonium hyperfine splitting \cite{FNAL}, as the estimates 
from relativistic \cite{FNAL,D234} and NRQCD \cite{HDT} actions 
are all substantially smaller than the experimental 
value of $(118\pm2)$~MeV.

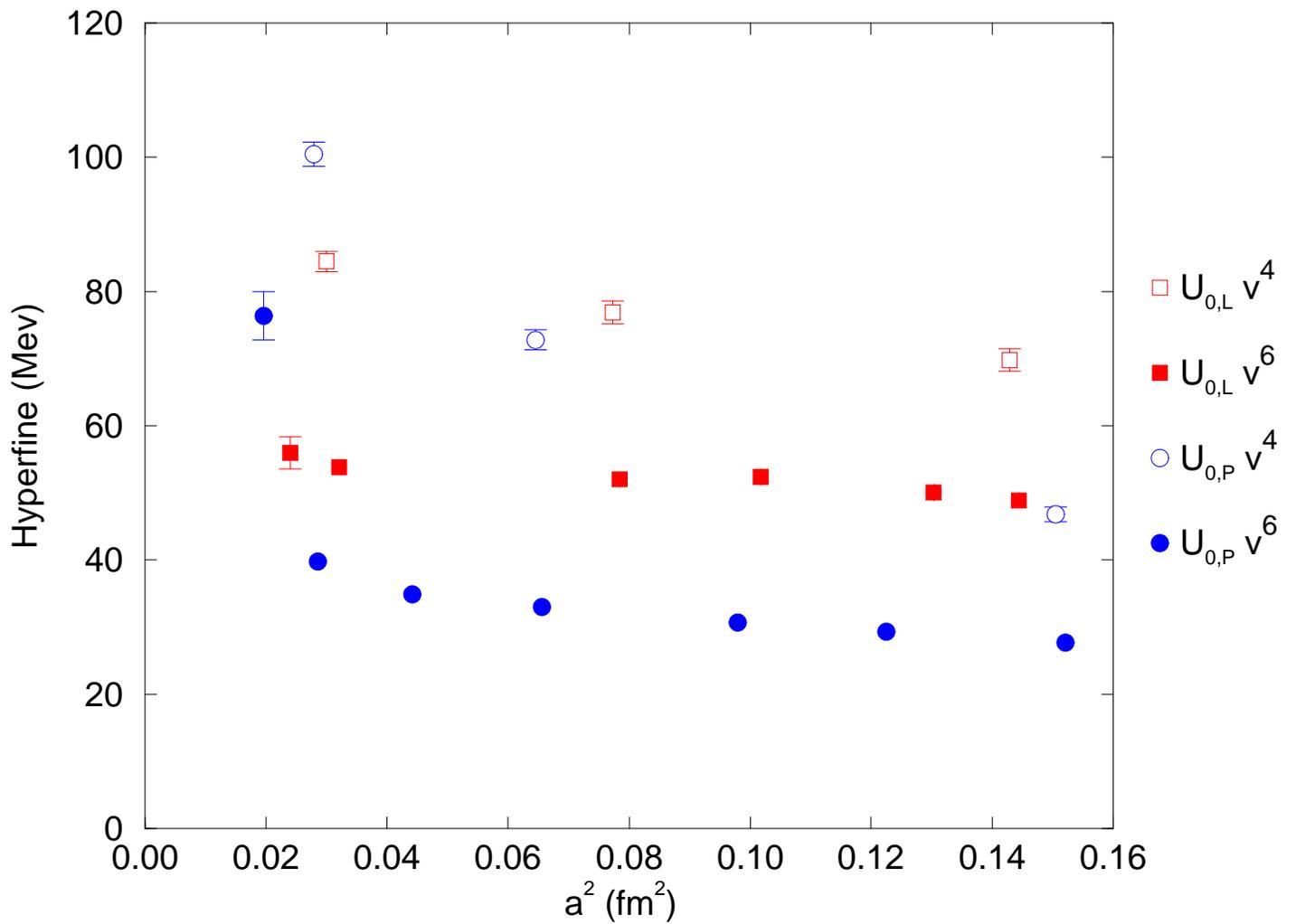
\begin{figure}
\caption{Hyperfine splittings for the $c\bar c$ system
versus lattice spacing squared.}
\label{FHcc}
\end{figure}

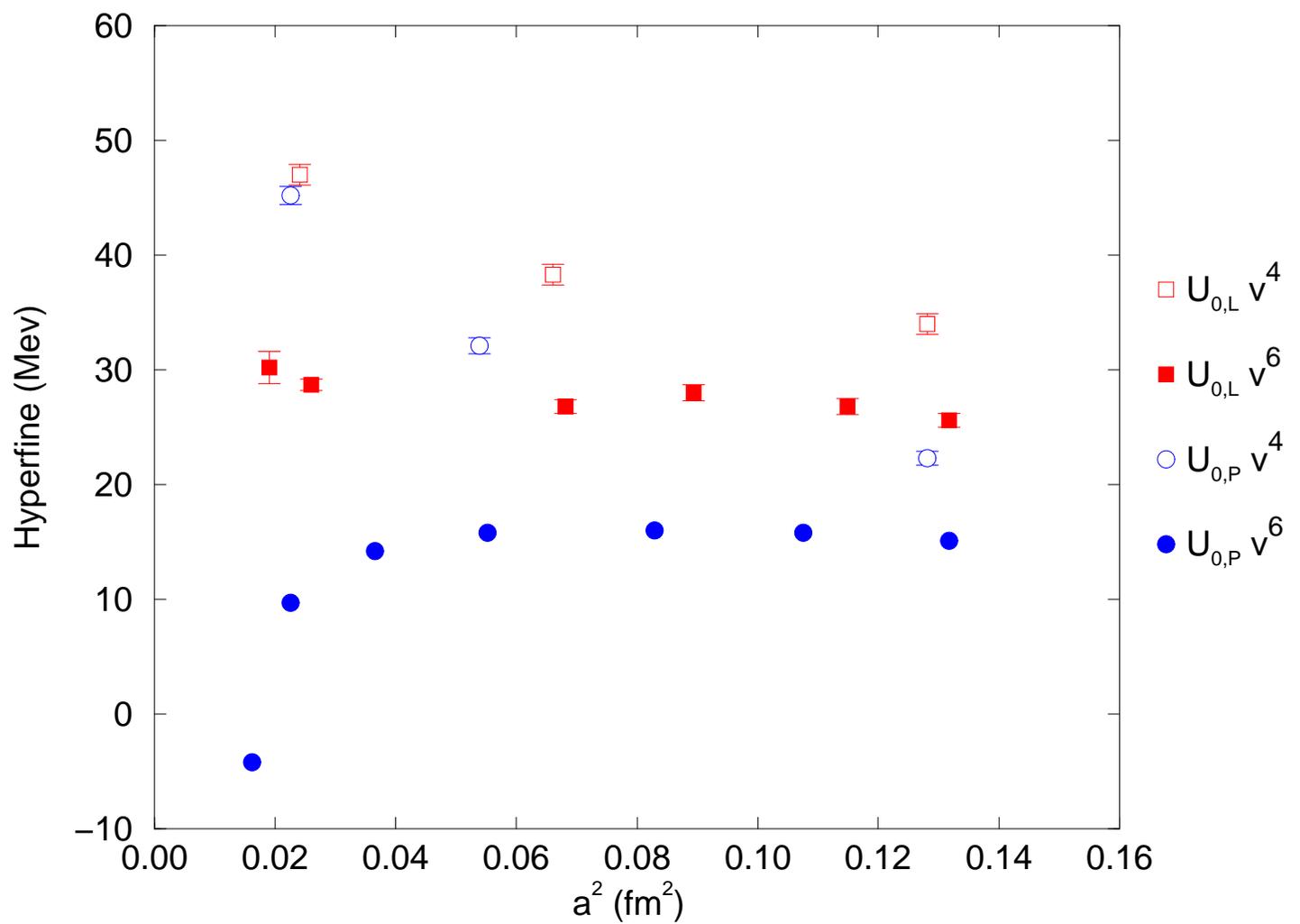
\begin{figure}
\caption{Hyperfine splittings for the $b\bar c$ system
versus lattice spacing squared.}
\label{FHbc}
\end{figure}

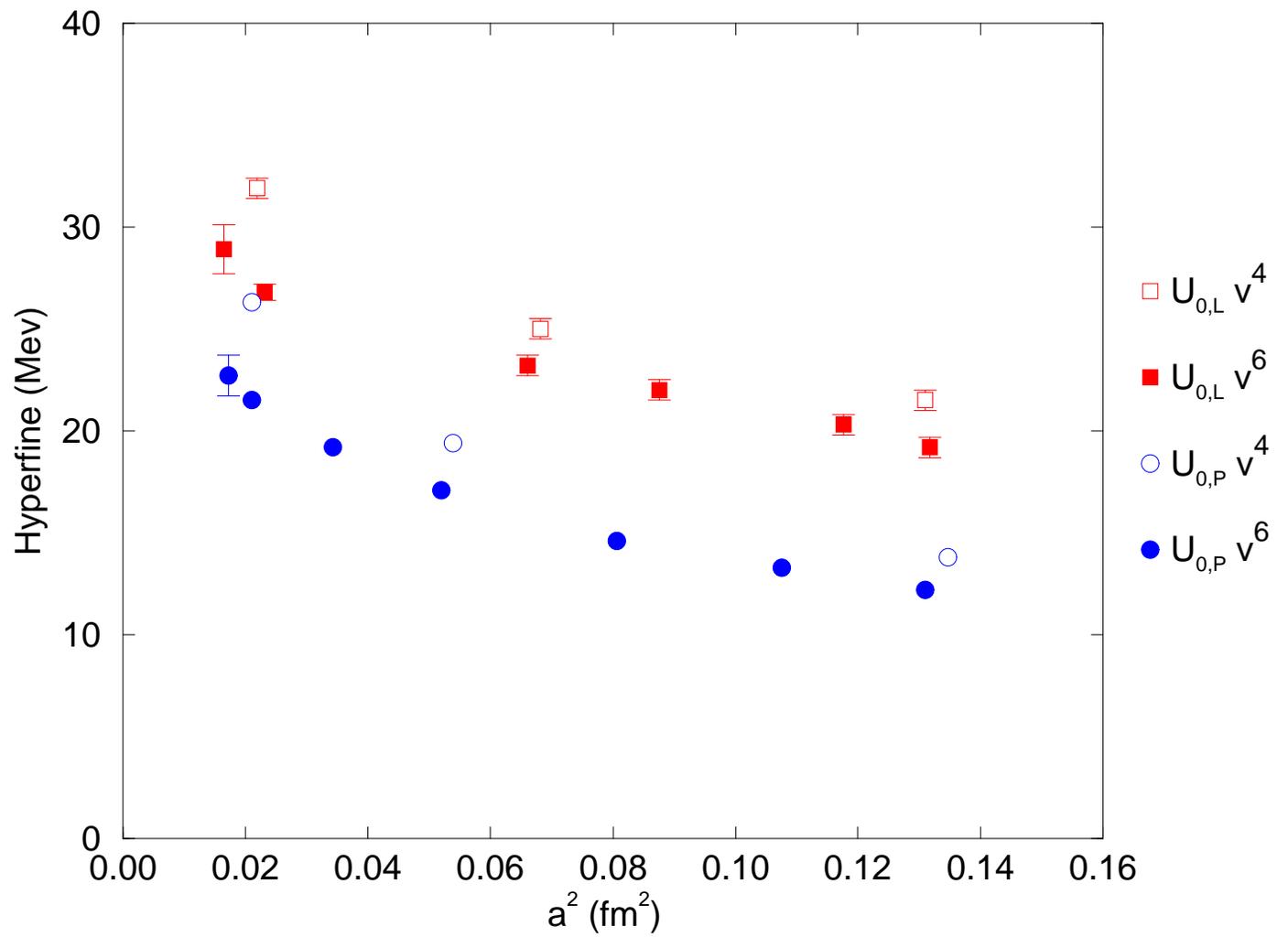
\begin{figure}
\caption{Hyperfine splittings for the $b\bar b$ system
versus lattice spacing squared.}
\label{FHbb}
\end{figure}

The splittings for the $c\bar c$, $b\bar c$, and the $b\bar b$ systems 
are plotted against lattice spacing squared in 
Figs. \ref{FHcc}, \ref{FHbc}, and \ref{FHbb} respectively. 
For each system the splittings obtained 
with $u_{0,L}$ and $u_{0,P}$, and at $O(v^4)$ and $O(v^6)$, are
plotted together in Figs. \ref{FHcc}--\ref{FHbb}.
We collect all results with $u_{0,L}$ in Fig.\ \ref{FHu0L}, and all 
results with $u_{0,P}$ in Fig.\ \ref{FHu0P}.

There are a number of very clear features in the data. To begin with, 
we note that the results with $u_{0,L}$ for the three quarkonium
systems show much smaller scaling violations than the results with 
$u_{0,P}$ (compare Figs. \ref{FHu0L} and \ref{FHu0P}).
The smallest scaling violations are in the results with 
$u_{0,L}$ at $O(v^6)$, which show remarkably little change 
as the lattice spacing is varied here by a factor of about 2.5.
The $b\bar b$ data show the largest scaling 
violations, as expected from the fact that this system should
have the smallest size of the three. 
This scaling analysis provides evidence that $u_{0,L}$ tadpole 
renormalization yields a more continuum-like action than does $u_{0,P}$. 
The improved scaling may also demonstrate indirectly that $O(v^6)$ 
corrections improve the matching of NRQCD to true QCD, as expected.

Perhaps the most striking feature of the data is the sharp drop in 
the $b\bar c$ splitting at smaller lattice spacings, when $u_{0,P}$ 
is used at $O(v^6)$ (see the filled circles in Fig.\ \ref{FHbc}). 
In fact, most of the $c$-quark data with $u_{0,P}$ show very large 
changes at the smallest lattice spacings. The $u_{0,L}$ data on the 
other hand exhibit a much smoother behavior. Note for example that the 
$O(v^6)$ $c\bar c$ splitting with $u_{0,P}$ lies well below the data 
with $u_{0,L}$, except at the smallest lattice spacings, where the 
$u_{0,P}$ data show a sharp upturn (compare the filled circles
and filled squares in Fig.\ \ref{FHcc}).

We interpret these features as possible indicators of a breakdown 
in the NRQCD effective action at smaller lattice spacings, when the bare 
quark mass in lattice units $aM_Q^0$ falls below about one. From 
Tables \ref{TbaremL} and \ref{TbaremP} we see that the bare $c$-quark 
mass for example turns out to be much larger when $u_{0,L}$ is used, 
compared to when $u_{0,P}$ is used on lattices with comparable 
spacings (compare $aM_c^0=0.65$ at $a_P \approx 0.14$~fm with
$aM_c^0=1.10$ at $a_L \approx 0.16$~fm). Although there is clearly a 
pathological behavior in the $c$-quark simulations with $u_{0,P}$ 
at smaller lattice spacings, we cannot reach definitive conclusions 
regarding the breakdown of the effective action without doing $u_{0,L}$ 
simulations at much smaller $aM_c^0$. This requires lattices with
much smaller spacings than we can realistically simulate.

Another key feature of these results is that the relativistic
corrections to the hyperfine splittings are smaller 
when the action is renormalized using $u_{0,L}$,
particularly for the $c\bar c$ and $b\bar c$ systems. 
For example, we find that the charmonium hyperfine splitting is 
reduced by about 30--40\% in going from $O(v^4)$ to $O(v^6)$ 
when using $u_{0,L}$, compared to a reduction of about 40--60\% 
when using $u_{0,P}$. Note that the $u_{0,P}$ estimate of the
relativistic corrections depends very strongly on the lattice 
spacing, increasing rapidly as $a$ decreases; this may be related to 
pathologies in the data at $aM_c^0 \alt 1$, discussed above.

The size of the corrrection to the hyperfine splitting obtained 
with $u_{0,L}$ is consistent with a naive estimate of 
$v^2_{c\bar c}$ from operator expectation values in 
NRQCD \cite{N1992}. Hence while the velocity 
expansion for charmonium is subject to large corrections, it may 
not be as unreliable as was suggested in Ref.\ \cite{HDT}, based 
on simulations with $u_{0,P}$ on fine lattices.

We note finally that it is reasonable to attempt to extrapolate
the $O(v^6)$ hyperfine splittings for $c\bar c$ and $b\bar c$
to zero lattice spacing, from the data on coarse lattices, where 
there is reasonably good scaling behavior (and where cutoff effects 
in the effective theory should be small, since $aM_Q^0 > 1$ 
in this region \cite{bbscaling}). However, the extrapolations 
in the $u_{0,L}$ and $u_{0,P}$ data, both of which exhibit
good scaling on coarse lattices, are clearly very different
(compare the filled circles with the filled squares 
in Fig. \ref{FHcc}, and in Fig. \ref{FHbc}). This suggests that 
some relevant operator coefficients $c_i$ in the NRQCD action
(Eqs. (\ref{H4}) and (\ref{H6})) receive significant $O(\alpha_s)$ 
corrections in one or both of the two tadpole schemes. This 
underlines the need to go beyond tree-level tadpole improvement 
in order to fully clarify the differences between
renormalization schemes.

\section{Summary}

We have presented new evidence that clearly favors tadpole 
renormalization using the mean-link in Landau gauge over the fourth 
root of the average plaquette. This includes a demonstration of 
much better scaling behavior of the hyperfine splittings in three 
quarkonium systems when $u_{0,L}$ is used, and a smaller size 
for spin-dependent relativistic corrections.
The results presented here also help to elucidate the structure
of the NRQCD effective action. In particular, we see signs of a 
breakdown in the NRQCD expansion when the bare quark mass falls 
below about one in lattice units, with pathological behavior clearly
visible in the $c$-quark systems with $u_{0,P}$ tadpoles. 
With $u_{0,L}$ on the other hand the bare quark masses turn out 
to be much larger than with $u_{0,P}$, resulting in a much
smoother behavior. We have in fact been unable to do realistic 
simulations on lattice fine enough to make 
$aM_c^0<1$ with $u_{0,L}$ tadpoles.
At the same time, these results also serve to underline the 
need to go beyond tree-level matching of improved actions, in
order to eliminate uncertainties due to uncalculated $O(\alpha_s)$
renormalizations \cite{Luscher,LepHDT}.

\acknowledgments

We are indebted to C.~T.~H. Davies, G.~P. Lepage, R. Lewis, 
and R.~M. Woloshyn for many helpful discussions and suggestions.
We also thank T. Manke for useful comments.
This work was supported in part by the 
Natural Sciences and Engineering Research Council of Canada.


\begin{table}
\begin{center}
\begin{tabular}{dcccc}
  $\beta_L$   
& $\langle \case13 \mbox{ReTr} \, U_\mu \rangle$
& $\langle \case13 \mbox{ReTr} \, U_{\mbox{pl}} \rangle^{1/4}$
& $a_{c\bar c}$ (fm)    & Volume \\
\hline
7.5  &  0.836   &  0.879   & 0.16  & $12^3 \times 16$ \\
7.4  &  0.829   &  0.875   & 0.18  & $10^3 \times 16$ \\
7.0  &  0.780   &  0.850   & 0.28  & $6^3 \times 10$  \\
6.85 &  0.763   &  0.840   & 0.32  & $6^3 \times 10$  \\
6.7  &  0.750   &  0.830   & 0.36  & $6^3 \times 10$  \\
6.6  &  0.743   &  0.825   & 0.38  & $6^3 \times 10$  \\
\end{tabular}
\end{center}
\caption{Simulation parameters using the Landau gauge mean-link 
to determine the tadpole renormalization $u_{0,L}$ 
(second column). The lattice spacing determined
from the charmonium system is given as a guide.}
\label{betaL}
\end{table}

\begin{table}
\begin{center}
\begin{tabular}{dcccc}
  $\beta_P$   
& $\langle \case13 \mbox{ReTr} \, U_\mu \rangle$
& $\langle \case13 \mbox{ReTr} \, U_{\mbox{pl}} \rangle^{1/4}$
& $a_{c\bar c}$ (fm)    & Volume  \\
\hline
7.3    &  0.837   &  0.878   & 0.14   & $12^3 \times 16$ \\
7.2    &  0.830   &  0.875   & 0.17   & $10^3 \times 16$ \\
7.0    &  0.812   &  0.864   & 0.21   & $8^3 \times 10$  \\
6.8    &  0.791   &  0.854   & 0.26   & $6^3 \times 10$  \\
6.6    &  0.771   &  0.841   & 0.31   & $6^3 \times 10$  \\
6.4    &  0.753   &  0.829   & 0.35   & $6^3 \times 10$  \\
6.25   &  0.741   &  0.821   & 0.39   & $6^3 \times 10$  \\
\end{tabular}
\end{center}
\caption{Simulation parameters using the average plaquette
to determine the tadpole renormalization $u_{0,P}$ (third column).}
\label{betaP}
\end{table}

\begin{table}
\begin{center}
\begin{tabular}{dcccccc}
  & \multicolumn{2}{c}{$c \bar c$}
  & \multicolumn{2}{c}{$b \bar c$}
  & \multicolumn{2}{c}{$b \bar b$} \\
$\beta_L$     
  & $a$ (fm)  & $aM_c^0[n]$   
  & $a$ (fm)  & $aM_b^0[n]$
  & $a$ (fm)  & $aM_b^0[n]$ \\
\hline
7.5  & 0.155(4) & 1.10[4]  & 0.138(4) & 3.20[2]  & 0.128(3) & 3.20[2] \\
7.4  & 0.179(2) & 1.20[4]  & 0.161(2) & 3.57[2]  & 0.152(2) & 3.57[2] \\
7.0  & 0.280(4) & 1.97[2]  & 0.261(4) & 6.10[2]  & 0.257(3) & 5.35[2] \\
6.85 & 0.319(5) & 2.25[2]  & 0.299(5) & 6.50[2]  & 0.296(4) & 5.90[2] \\
6.7  & 0.361(6) & 2.50[2]  & 0.339(6) & 7.20[2]  & 0.343(6) & 6.35[2] \\
6.6  & 0.380(7) & 2.67[2]  & 0.363(7) & 7.50[2]  & 0.363(7) & 6.66[2] \\
\end{tabular}
\end{center}
\caption{Lattice spacings and bare quark masses for the three
quarkonium systems at $O(v^6)$, using Landau gauge mean-link 
tadpoles $u_{0,L}$; the stability parameter $n$ for each mass 
is given in square brackets.}
\label{TbaremL}
\end{table}

\begin{table}
\begin{center}
\begin{tabular}{dcccccc}
  & \multicolumn{2}{c}{$c \bar c$}
  & \multicolumn{2}{c}{$b \bar c$}
  & \multicolumn{2}{c}{$b \bar b$} \\
$\beta_P$     
  & $a$ (fm)  & $aM_c^0[n]$   
  & $a$ (fm)  & $aM_b^0[n]$
  & $a$ (fm)  & $aM_b^0[n]$ \\
\hline
7.3  & 0.140(4) & 0.65[8]  & 0.131(4) & 2.87[2]  & 0.127(4) & 2.87[2] \\
7.2  & 0.169(2) & 0.83[4]  & 0.150(2) & 3.20[2]  & 0.145(2) & 3.20[2] \\
7.0  & 0.210(2) & 1.10[4]  & 0.191(2) & 4.10[2]  & 0.185(2) & 3.95[2] \\
6.8  & 0.256(3) & 1.43[3]  & 0.235(3) & 4.98[2]  & 0.228(3) & 4.53[2] \\
6.6  & 0.313(4) & 1.80[3]  & 0.288(4) & 5.83[2]  & 0.284(4) & 5.23[2] \\
6.4  & 0.350(6) & 2.15[2]  & 0.328(6) & 6.45[2]  & 0.328(6) & 5.60[2] \\
6.25 & 0.390(6) & 2.41[2]  & 0.363(7) & 6.85[2]  & 0.362(6) & 5.99[2] \\
\end{tabular}
\end{center}
\caption{Lattice spacings and bare quark masses for the three
quarkonium systems at $O(v^6)$ using average plauqette tadpoles 
$u_{0,P}$.}
\label{TbaremP}
\end{table}

\begin{table}
\begin{center}
\begin{tabular}{ccccc}
$t_{\rm min}/t_{\rm max}$  
& ${}^1P_1$ & ${}^3S_1$    & ${}^1S_0$    & ${}^3S_1 - {}^1S_0$   \\
\hline
2/16  & 0.627(9)      & 0.306(2)  & 0.285(1)  & 0.0210(3)    \\
3/16  & 0.621(10)     & 0.305(2)  & 0.284(1)  & 0.0212(3)    \\
4/16  & 0.618(12)     & 0.305(2)  & 0.284(1)  & 0.0212(4)    \\
5/16  & 0.615(14)     & 0.305(2)  & 0.284(2)  & 0.0210(4)    \\
6/16  & 0.612(17)     & 0.304(2)  & 0.283(2)  & 0.0209(4)    \\
7/16  & 0.606(20)     & 0.304(2)  & 0.283(2)  & 0.0207(5)    \\
8/16  & 0.598(24)     & 0.303(2)  & 0.283(2)  & 0.0205(5)    \\
9/16  & 0.586(30)     & 0.303(2)  & 0.283(2)  & 0.0205(6)    \\
10/16 & 0.591(39)     & 0.304(2)  & 0.283(2)  & 0.0204(7)    \\
\end{tabular}
\end{center}
\caption{Examples of fits to the $O(v^6)$ $b\bar c$ spectra at 
$\beta_L=7.5$ ($a \approx .14$~fm).}
\label{TbcL}
\end{table}

\begin{table}
\begin{center}
\begin{tabular}{ccccc}
$t_{\rm min}/t_{\rm max}$  
& ${}^1P_1$   & ${}^3S_1$    & ${}^1S_0$   & ${}^3S_1 - {}^1S_0$   \\
\hline
2/16  &  0.879(9)    &  0.578(2)  &  0.581(2)  &  -0.0029(3)   \\
3/16  &  0.874(10)   &  0.577(2)  &  0.579(2)  &  -0.0028(3)   \\
4/16  &  0.872(12)   &  0.577(2)  &  0.579(2)  &  -0.0027(3)   \\
5/16  &  0.870(14)   &  0.577(2)  &  0.580(2)  &  -0.0026(4)   \\
6/16  &  0.871(17)   &  0.577(2)  &  0.580(2)  &  -0.0026(4)   \\
7/16  &  0.877(21)   &  0.579(2)  &  0.581(2)  &  -0.0026(5)   \\
8/16  &  0.885(25)   &  0.579(2)  &  0.582(2)  &  -0.0027(5)   \\
\end{tabular}
\end{center}
\caption{Examples of fits to the $O(v^6)$ $b\bar c$ spectra
at $\beta_P=7.3$ ($a \approx .13$~fm).}
\label{TbcP}
\end{table}

\begin{table}
\begin{center}
\begin{tabular}{dcccc}
$\beta_L$
  & ${}^1P_1$   & ${}^3S_1$    & ${}^1S_0$   & ${}^3S_1 - {}^1S_0$   \\
\hline
7.5  & 0.62(1)  & 0.305(2)  & 0.284(1)  & 0.0212(3)  \\
7.4  & 0.70(1)  & 0.330(1)  & 0.306(1)  & 0.0235(2)  \\
7.0  & 0.95(1)  & 0.346(1)  & 0.310(1)  & 0.0356(3)  \\
6.85 & 1.01(1)  & 0.329(1)  & 0.287(1)  & 0.0425(3)  \\
6.7  & 1.08(2)  & 0.308(1)  & 0.261(1)  & 0.0460(4)  \\
6.6  & 1.12(2)  & 0.292(1)  & 0.245(1)  & 0.0471(3)  \\
\end{tabular}
\end{center}
\caption{Final fit results for the dimensionless energies
for the $O(v^6)$ $b\bar c$ spectra using $u_{0,L}$.}
\label{TbcfitL}
\end{table}

\begin{table}
\begin{center}
\begin{tabular}{dcccc}
$\beta_P$
  & ${}^1P_1$   & ${}^3S_1$    & ${}^1S_0$   & ${}^3S_1 - {}^1S_0$   \\
\hline
7.3  & 0.87(1)  & 0.577(2)  & 0.580(2)  & -0.0027(3)  \\
7.2  & 0.98(1)  & 0.637(1)  & 0.629(1)  &  0.0074(1)  \\
7.0  & 1.12(1)  & 0.681(1)  & 0.666(1)  &  0.0138(1)  \\
6.8  & 1.25(1)  & 0.706(1)  & 0.687(1)  &  0.0189(2)  \\
6.6  & 1.36(1)  & 0.699(1)  & 0.676(1)  &  0.0233(2)  \\
6.4  & 1.44(2)  & 0.681(1)  & 0.654(1)  &  0.0263(2)  \\
6.25 & 1.48(2)  & 0.648(1)  & 0.621(1)  &  0.0277(2)  \\
\end{tabular}
\end{center}
\caption{Final fit results for the dimensionless energies
for the $O(v^6)$ $b\bar c$ spectra using $u_{0,P}$.}
\label{TbcfitP}
\end{table}

\begin{table}
\begin{center}
\begin{tabular}{ddddddd}
 & \multicolumn{3}{c}{$O(v^6)$}
 & \multicolumn{3}{c}{$O(v^4)$} \\
$\beta_L$ & $c\bar c$  & $b\bar c$  & $b\bar b$
          & $c\bar c$  & $b\bar c$  & $b\bar b$ \\
\hline
7.5  & 56.0(24) & 30.2(14) & 28.9(12) &  &  &  \\
7.4  & 53.8(9)  & 28.7(5)  & 26.8(4)  & 84.5(15)  & 47.0(9) & 31.9(5) \\
7.0  & 52.0(11) & 26.8(6)  & 23.2(5)  & 76.9(17)  & 38.3(9) & 25.0(5) \\
6.85 & 52.4(12) & 28.0(7)  & 22.0(5)  &  &  &  \\
6.7  & 50.1(12) & 26.8(7)  & 20.3(5)  &  &  &  \\
6.6  & 48.9(11) & 25.6(6)  & 19.2(5)  & 69.8(17)  & 34.0(9) & 21.5(5) \\
\end{tabular}
\end{center}
\caption{Final fit results for hyperfine splittings in MeV
for the three quarkonium systems, with $u_{0,L}$ tadpole
renormalization.}
\label{THL}
\end{table}

\begin{table}
\begin{center}
\begin{tabular}{ddddddd}
 & \multicolumn{3}{c}{$O(v^6)$}
 & \multicolumn{3}{c}{$O(v^4)$} \\
$\beta_P$   & $c\bar c$  & $b\bar c$  & $b\bar b$
            & $c\bar c$  & $b\bar c$  & $b\bar b$ \\
\hline
7.3  & 76.4(36) & -4.2(3)  & 22.7(10) &  &  &  \\
7.2  & 39.8(7)  &  9.7(2)  & 21.5(4)  & 100.4(18)  & 45.2(8) & 26.3(4) \\
7.0  & 34.9(6)  & 14.2(3)  & 19.2(4)  &  &  &  \\
6.8  & 33.0(7)  & 15.8(3)  & 17.1(3)  &  72.8(15)  & 32.1(7) & 19.4(4) \\
6.6  & 30.7(6)  & 16.0(4)  & 14.6(3)  &  &  &  \\
6.4  & 29.3(7)  & 15.8(4)  & 13.3(3)  &  &  &  \\
6.25 & 27.7(6)  & 15.1(4)  & 12.2(3)  &  46.8(11)  & 22.3(6) & 13.8(3) \\
\end{tabular}
\end{center}
\caption{Final fit results for hyperfine splittings in MeV
for the three quarkonium systems, with $u_{0,P}$ tadpole
renormalization.}
\label{THP}
\end{table}


\begin{figure}
\caption{Effective mass plot for $O(v^6)$ $b\bar c$ spectra
at $\beta_L=7.5$ ($a \approx 0.14$~fm): 
(a) ${}^1P_1$ state ($\Box$) and ${}^1S_0$ state ($\circ$);
(b) hyperfine splitting.}
\label{FbcL}
\end{figure}

\begin{figure}
\caption{Effective mass plot for $O(v^6)$ $b\bar c$ spectra
at $\beta_P=7.3$ ($a \approx 0.13$~fm):
(a) ${}^1P_1$ state ($\Box$) and ${}^1S_0$ state ($\circ$);
(b) hyperfine splitting.}
\label{FbcP}
\end{figure}

\begin{figure}
\caption{Hyperfine splittings with $u_{0,L}$ versus lattice
spacing squared.}
\label{FHu0L}
\end{figure}

\begin{figure}
\caption{Hyperfine splittings with $u_{0,P}$ versus lattice
spacing squared.}
\label{FHu0P}
\end{figure}


\begin{references}
\bibitem{NoteSmear} 
The smearing function $\Gamma(\vec x)$ used here is identical 
to the one used in Ref.\ \cite{HDT}. Note that Eq.\ (22) for
$\Gamma(\vec x)$ in Ref.\ \cite{HDT} is incorrect, and should read 
as given here in Eq.\ (\ref{Gamma}).
\end{references}
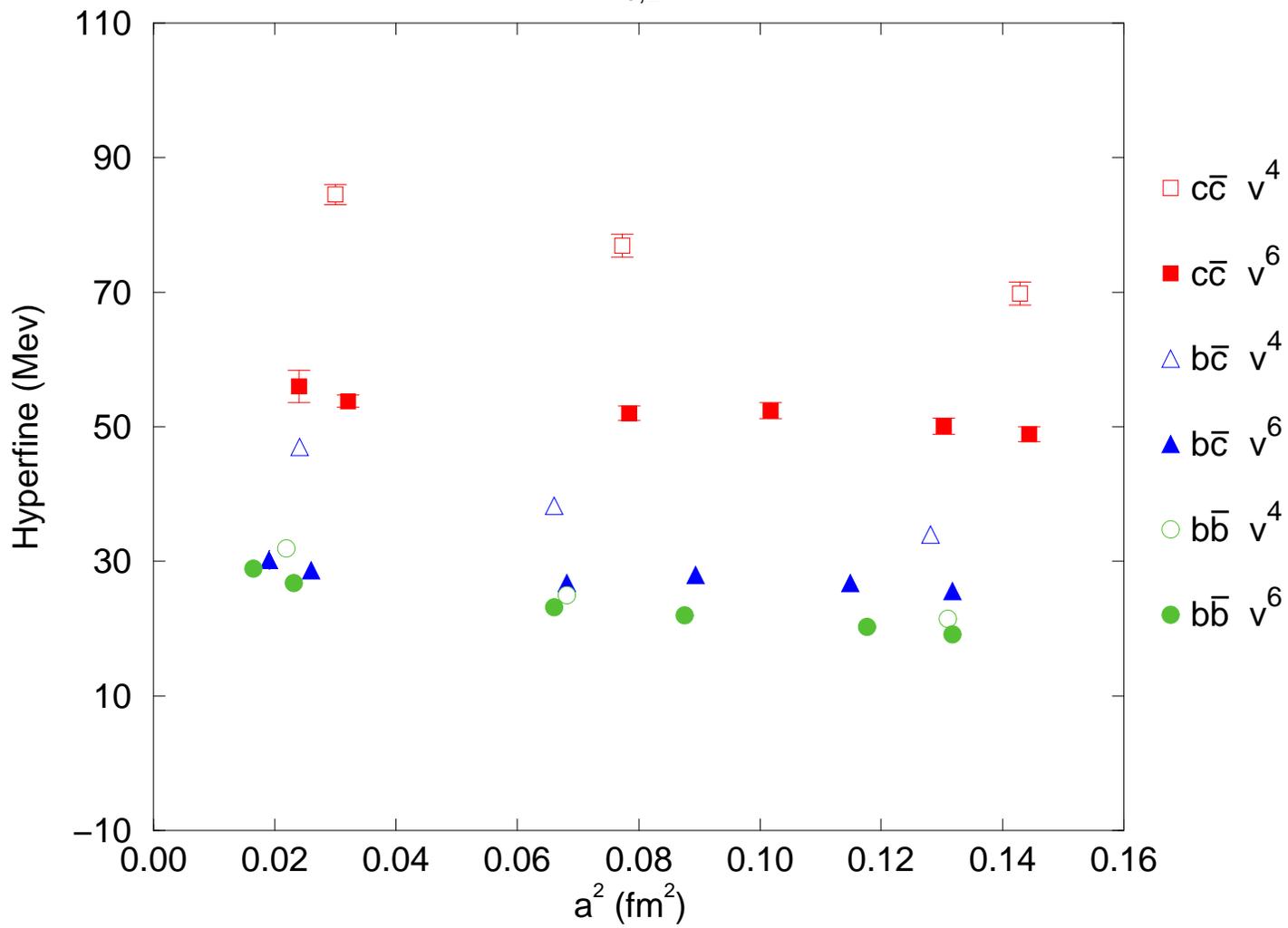
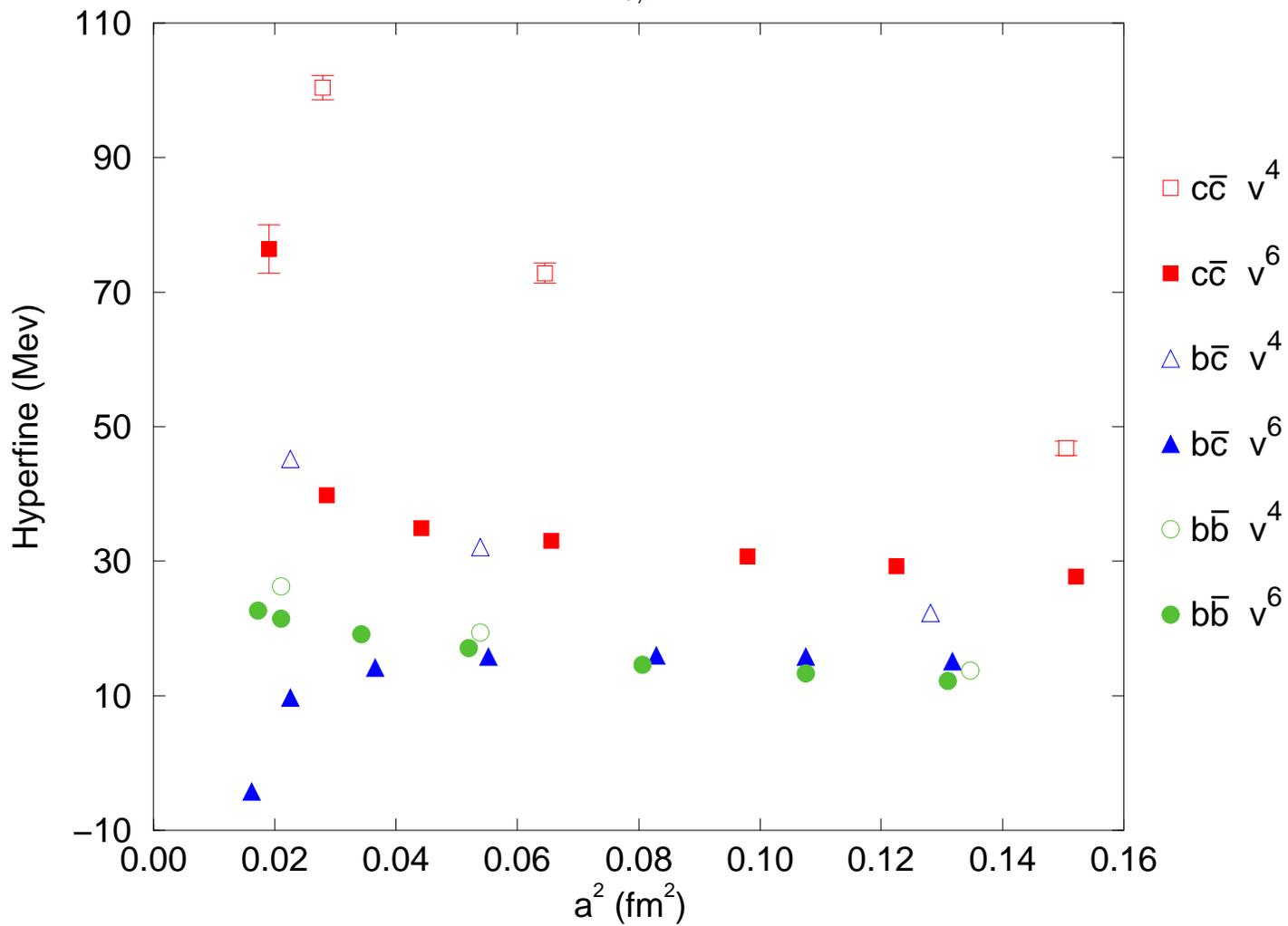
\end{document}